\documentclass[a4paper,twocolumn]{article}

\usepackage[utf8]{inputenc}
\usepackage[T1]{fontenc}
\usepackage{float}
\usepackage{amsmath,amssymb}
\usepackage{mathrsfs}
\usepackage{theorem}
\usepackage{graphicx}
\usepackage{color}
\usepackage{wasysym}
\usepackage{hyperref}

\definecolor{blue}{rgb}{0,0,1}
\definecolor{green}{rgb}{0,0.65,0.5}
\definecolor{verde}{rgb}{0.,.5,0.4}
\definecolor{marron}{rgb}{0.7,0.2,0.1}
\definecolor{red}{rgb}{1,0,0}
\definecolor{vio}{rgb}{0.66,0,1}
\definecolor{viob}{rgb}{0.6,0.3,0.8}
\definecolor{ama}{rgb}{1,1,0}

\newcommand{\bc}{\begin{center}}
\newcommand{\ec}{\end{center}}
\newcommand{\be}{\nopagebreak[3]\begin{equation}}
\newcommand{\ee}{\end{equation}}
\newcommand{\ba}{\nopagebreak[3]\begin{eqnarray}}
\newcommand{\ea}{\end{eqnarray}}

{\theorembodyfont{\sffamily} }
{\theorembodyfont{\sffamily} }
{\theorembodyfont{\sffamily} }

\begin{document}

\title{\bf 
Constructing balanced equations of motion for
particles in general relativity: \\
the null gauge case
}

\author{
Emanuel Gallo
and
Osvaldo M. Moreschi
\\
{\rm \small Facultad de Matemática Astronomía, Física y Computación (FaMAF),} \\
{\rm \small Universidad Nacional de C\'{o}rdoba,}\\
{\rm \small Instituto de F\'\i{}sica Enrique Gaviola (IFEG), CONICET,}\\
{\rm \small Ciudad Universitaria,(5000) C\'{o}rdoba, Argentina.}
}

\maketitle

\begin{abstract}
	
We present a convenient null gauge for the construction of the
balanced equations of motion.

This null gauge has the property that
the asymptotic structure is intimately related to
the interior one; in particular there is a strong
connexion between the field equation and the
balanced equations of motion.

We present the balanced equations of motion in second order
of the acceleration. We solve the required components
of the field equation at their respective required orders,
$G^2$ and $G^3$.

We indicate how 
this approach can be extended to higher orders.

\end{abstract}



\section{Introduction}

With this new era of gravitational wave 
observations\cite{Abbott:2016blz,TheLIGOScientific:2016wfe,Abbott:2016nmj,Abbott:2017vtc,
	Abbott:2017oio,TheLIGOScientific:2017qsa}
coming from binary black holes and binary neutron stars,
there also comes the need to have at hand the most convenient models
of the physical system, to be used at every stage of their dynamics.
It has been said that:
``it is infeasible to use the NR simulations directly as search 
templates''\cite{Abadie:2011kd}.
Therefore researchers\cite{Abadie:2011kd} use ``phenomenological waveforms''.

We would like to contribute with models that are
useful for the description of the dynamics of coalescence binary systems
and their specific relation with the gravitational emitted radiation.
Our work concentrates in the construction of equations of motion
of compact objects, subjected to the back reaction due
to their emission of gravitational radiation.
	
This is a sequel of references \cite{Gallo:2016hpy} and \cite{Gallo:2017yys}
where we have presented the general
necessary framework to construct balanced equations of motion for particle
in general relativistic theories, and the specific application to the harmonic gauge. 
In this work we present this framework applied to the case of general relativity
in the null gauge.

As in the previous work,
 we have in mind is a bound binary gravitationally isolated
interactive system.
When considering an isolated compact object, 
it induces one to represent it by an asymptotically flat spacetime.
Then, in the asymptotic region one can always write the metric as
\begin{equation}\label{eq:asymp1}
g = \eta_{\text{asy}} + h_{\text{asy}} 
;
\end{equation}
where $\eta_{\text{asy}}$ is a flat metric associated to inertial frames in the asymptotic
region
and $h_{\text{asy}}$ the tensor where all the physical information is encoded.
But, as we have already noted,
there are as many flat metrics $\eta_{\text{asy}}$ as there are BMS\cite{Sachs62,Moreschi86} 
proper supertranslation generators.
We have explained in \cite{Gallo:2016hpy} the difficulties in finding
appropriate rest frames, and how to solve these issues;
making use of supertranslation free definition of intrinsic 
angular momentum\cite{Moreschi04,Gallo:2014jda}.
The root of all obstacles is the appearance of gravitational radiation\cite{Moreschi88,Moreschi98,Dain00'}.
Because of this when studying the dynamics of compact objects
we take into account the back reaction due to gravitational radiation
as our starting point.

Whenever necessary we will assign a label $A$ to the particle under consideration,
and use $B$ to denote the rest of the system, that is thought to be
the other particle, in the binary situation.

We adopt here the viewpoint explained in  \cite{Gallo:2016hpy}
in which we assume there exists an exact metric
$\mathsf{g}$  that corresponds to an isolated 
binary 	system of compact objects;
which it can be decomposed in a form:
\begin{equation}
\mathsf{g}  = \eta + \mathsf{h}_A + \mathsf{h}_B + \mathsf{h}_{AB} ;
\end{equation} 
where $\eta$ is a flat metric,
$ \mathsf{h}_A$ is proportional to a parameter $M_A$, that one can think is some kind
of measure of the mass of system $A$,
similarly $ \mathsf{h}_B$ is proportional to a parameter $M_B$,
and $ \mathsf{h}_{AB}$ is proportional to both parameters.
To study the gravitational radiation emitted by the motion of
particle $A$, 
we model the asymptotic structure of a sub-metric 
\begin{equation}\label{eq:gA}
g_A = \eta + h_A ;
\end{equation}
and to describe the rest of the system, we use
a sub-metric
\begin{equation}\label{eq:gB}
g_B = \eta + h_B .
\end{equation}
For more details see article \cite{Gallo:2016hpy},
where it is also explained that
the appropriate choice of the flat metric $\eta$ 
should be related to a local notion of center of mass frame.

Although we will be studying the dynamics of system $A$, to simplify the 
notation we will avoid using a subindex $A$, whenever possible.

We present here a model for a compact object, treated as a particle on
an appropriately chosen flat background. The idea one has in mind is to 
apply this construction to a binary system, so that each of the compact
objects will be treated likewise.
The flat background metric by construction will share the same asymptotic
region as the full metric of the spacetime: so that one of the inertial
system at infinity would be related to this flat global metric.
The model, for each monopole, will solve the field equations at appropriate orders
and by construction, the dynamics will balance the amount of gravitational
radiation emitted due to the acceleration of the body.

Although in paper \cite{Gallo:2016hpy} we have presented several delicate
issues that one has to consider
when constructing a balanced equations of motion,
here we present a model that is still simple and so 
it has the advantage that one should
be able to compute without recurring to supercomputers.
It is our intention to provide with this model a method for the calculation
of observational waveforms, in a wide range of astrophysical
parameters.
In future works we will apply this model to specific observations.

In section \ref{sec:interior} we describe the interior structure of 
the null gauge model.
Section \ref{sec:asympt} is devoted to the presentation of the main
concepts that arise in the neighborhood of future null infinity of an
asymptotically flat spacetime.
In section \ref{sec:mono} we present the null gauge model from
the monopole structure of a general asymptotically flat spacetime.
Our main result is presented in section \ref{sec:dynamics},
where the balanced equations of motion for the null gauge
is calculated up to second order in the accelerations.
The last section contains final comments on our work.

\section{The interior problem in terms of the null gauge}\label{sec:interior}

\subsection{The notion of a particle over a flat background} 

Let us consider a massive point particle with mass
$M$ describing, in a flat space-time $(M^0,\eta_{ab})$, a curve $C$
which in a Cartesian coordinate system $x^a$ reads
\begin{equation}\label{eq:trajectoria}
 x^\mu =z^\mu(\tau_0),
\end{equation}
with $\tau_0$ meaning the proper time of the particle along $C$.

{The unit tangent vector to $C$, with respect to the flat background metric is}
\begin{equation}\label{eq:4-velocity}
\mathbf{v}^\mu =\frac{dz^\mu}{d\tau_0},
\end{equation} 
that is, $\eta(\mathbf{v} , \mathbf{v})=1$.
For each point $p=C(\tau_0)$ let $\mathfrak{C}_p$ 
denote the future null cone with vertex at $p$. 
If we call $x^\mu_p$
the Minkowskian coordinates of
an arbitrary
point on the cone $\mathfrak{C}_p$, then we can define
the retarded radial distance from the point $p$  by
\begin{equation}\label{eq:retardedr}
r = \mathbf{v}_\mu \left(x^\mu_P-z^\mu(\tau_0) \right).
\end{equation}

Below we will introduce the four velocity vector $v^a$, proportional to
$\mathbf{v}^a$, but normalized with respect to the metric $g_B$.

\subsection{The null gauge near a particle over a non-flat background case}

Let us now consider the notion of a particle but in the context of a
smooth non-flat background.

Given our background metric, one can always construct, in a neighbourhood of the curve
$C$ the null surfaces formed by the future directed null geodesic,
emanating from points in $C$; that can be labeled by the null function $u$.

Using the null polar coordinate system $(x^{0},x^{1},x^{2},x^{3})=\left( u,r,%
(\zeta +\overline{\zeta}),\frac{1}{i}(\zeta -\overline{\zeta})\right)$,
in a neighborhood $C$,
one can express a null tetrad as:
\begin{equation}\label{eq:ellinitial}
\ell _{a}=\left( du\right)_{a} ,
\end{equation}
\begin{equation}
\ell ^{a}=\left( \frac{\partial }{\partial \,r}\right) ^{a} ,
\label{dos}
\end{equation}
\begin{equation}
m^{a}=\xi ^j\left( \frac{\partial }{\partial x^j}\right) ^{a} ,
\label{eq:vecm}
\end{equation}
\begin{equation}
\overline{m}^{a}=\overline{\xi}^j\left( \frac{\partial }{\partial x^j}\right) ^{a} ,
\label{tres}
\end{equation}
\begin{equation}\label{eq:vecn}
n^{a}=\,\left(\frac{\partial}{\partial \,u} \right)^{a}
+ \,U\,\left( \frac{\partial }{\partial \,r}\right)^{a}
+ X^j\,\left(\frac{\partial }{\partial \,x^j}\right)^{a} ;
\end{equation}
with $j=2,3$, and we use $a,b, \cdots$ as abstract indices.

This null tetrad satisfies
\begin{equation}
 g_{ab}\;l ^{a}\;n^{b} =1 ,
\end{equation}
and
\begin{equation}
 g_{ab}\;m^{a}\;\bar{m}^{b}= -1 ,
\end{equation}
and all other scalar products are zero.

Several useful expressions describing the null gauge we are using
can be found in reference \cite{Moreschi87}.

In particular for Minkowski spacetime the components of the null tetrad are
given by:
\begin{equation} \label{eq:U_M}
U = \frac{\dot V_\eta}{V_\eta} r -\frac{1}{2},
\end{equation}
\begin{equation} \label{eq:xi2_M}
\xi^2 = \frac{\xi^2_0}{r} = \frac{\sqrt{2} P_0 V_\eta }{r}, 
\end{equation}
\begin{equation} \label{eq:xi3_M}
\xi^3 = \frac{\xi^3_0}{r} = - i \frac{\sqrt{2} P_0 V_\eta }{r}, 
\end{equation}
\begin{equation} \label{eq:X_M}
X^j = 0; 
\end{equation}
where a  dot means $\partial /\partial u$,
 $P_{0}=(1+\zeta \bar\zeta)/2$, and $V_\eta$ is given by the following expression 
\begin{equation} \label{eq:VM}
V_\eta = \hat l^a \mathbf{v}^b \eta_{ab},
\end{equation}
where $(\eta_{\mu\nu})=\text{diag}(1,-1,-1-1)$, $\mathbf{v}^\mu=\mathbf{v}^\mu(u)$ is the four velocity
in Minkowski space-time, which depends only on $u$ and satisfies the normalization 
\begin{equation}
\mathbf{v}^a \,\mathbf{v}^b \, \eta_{ab}=1, 
\end{equation}
and the null vector $\hat l^a$ is given by
\begin{equation} \label{eq:lhat}
\begin{split}
(\hat l^\nu) 
=&
\bigg( 1, \sin(\theta)\cos(\phi), \sin(\theta)\sin(\phi),\cos(\theta) \bigg) \\
=&
\bigg( 1,\frac{\zeta +\bar \zeta }{
	1+\zeta \bar{\zeta }},\frac{\zeta -\bar{\zeta }}{i(1+\zeta 
	\bar{\zeta )}},\frac{\zeta \bar{\zeta }-1}{1+\zeta \bar{\zeta 
	}}\bigg) \\
=&
\bigg(
\sqrt{4\pi} Y_{00} , 
- \sqrt{\frac{2\pi}{3}} \big(Y_{1,1} - Y_{1,-1} \big) , \\
& \quad i \sqrt{\frac{2\pi}{3}} \big(Y_{1,1} + Y_{1,-1} \big) ,
 \sqrt{\frac{4\pi}{3}} Y_{1,0}
\bigg)
.
\end{split}
\end{equation}
We distinguish between the abstract indices $a,b, \cdots$ and the numeric 
indices $\mu, \nu, \cdots =0,1,2,3$; since in some cases it is convenient
to remark the tensorial or vectorial character of the equations, and
in others it is convenient to remark how to make the calculations
in a particular frame of reference.
The relation between these null coordinates and the Cartesian ones $y^\mu$  is the following
\begin{equation}\label{eq:ymu}
y^\mu = z^\mu(u)+rl^\mu(u,x^2,x^3) ;
\end{equation}
where $z^\mu(u)$ is a world line with unit tangent vector $u^\mu$,
and the Cartesian components of the vector $l$ are 
\begin{equation}\label{eq:ele}
l^\mu=\frac{\hat l^\mu} {V_\eta}. 
\end{equation}
Let us note that
\begin{equation}
l^\mu \,\mathbf{v}^\nu \, \eta_{\mu \nu} = 1 .
\end{equation}

The complex null vectors $\hat m$ and $\hat{\bar m}$, defined as
\begin{equation}\label{eq:hatm}
\hat m 
=  \xi^i_0 \frac{\partial}{\partial x^i}
=  \xi^2_0 \frac{\partial}{\partial x^2} + \xi^3_0 \frac{\partial}{\partial x^3} ,
\end{equation}
are the natural basis for the unit sphere $r=1$.
So we will use this as basis vectors in the subspace of the spheres $u=$constant,
$r=$constant.

\section{General asymptotically flat spacetimes}\label{sec:asympt}

\subsection{The leading order behaviour of an adapted null tetrad}

Let now $u$ denote a null hypersurface that contains future directed null geodesics
that reach future null infinity.
Then we can use the same null tetrad prescription, that we used before, but now
adapted to this null congruence.

The components $\xi^{i}$, $U$ and $X^{i}$ have the asymptotic expansion:
\begin{equation}
\xi ^{2}=\frac{\xi _{0}^{2}}{r} + O\left(\frac{1}{r^2}\right),
\qquad \xi ^{3}=\frac{\xi_{0}^{3}}{r} + O\left(\frac{1}{r^2}\right) , 
\end{equation}
with
\begin{equation}\label{eq:xileading}
\xi _{0}^{2}=\sqrt{2}P_{0}\;V,\qquad \xi _{0}^{3}=-i\xi _{0}^{2},
\end{equation}
where $V=V(u,\zeta,\bar\zeta)$ and
the square of $P_0 = \frac{(1+\zeta \bar\zeta)}{2}$ is the conformal factor 
of the unit sphere;
\begin{equation}
U=rU_{00}+U_{0}+\frac{U_{1}}{r} +  O\left(\frac{1}{r^2}\right), 
\end{equation}
where
\begin{equation}\label{eq:uceros}
U_{00}=\frac{\dot{V}}{V},\quad U_{0}=-\frac{1}{2}K_{V},\quad U_{1}=-\frac{%
\Psi _{2}^{0}+\overline{\Psi}_{2}^{0}}{2}, 
\end{equation}
where $K_V$ is the  curvature of the 2-metric
\begin{equation}\label{eq:desphere}
ds_V^{2}=\frac{1}{V^2 \, P_0^{2}}\;d\zeta \;d\bar{\zeta} ;
\end{equation}
where the regular conformal metric restricted to scri is precisely
$\tilde g \mid_{\cal I^+} = -  ds_V^2$. In terms of the edth operator
$\eth_V$ 
of the sphere (\ref{eq:desphere}) the curvature $K_V$ is given by
\begin{equation}\label{eq:kv}
K_{V}=\frac{2}{V}~\overline{\eth }_{V}\eth _{V}\, V-\frac{2}{V^{2}}~\eth _{V}V~%
\overline{\eth }_{V}V+V^{2} .
\end{equation}
Finally, the other components of the vector $n^a$ have the asymptotic form
\begin{equation}
 X^{2}= O\left(\frac{1}{r^2}\right) ,
\qquad X^{3}= O\left(\frac{1}{r^2}\right)  .
\end{equation}

One can see that the previous expressions can be consider as a subset of
the present equations; since the first are also expressing an asymptotically flat
spacetime.

\subsection{The total momentum and flux}
Given any section $S$ at future null infinity, the total momentum
{of a generic spacetime}, in terms of an
inertial (Bondi) frame\cite{Moreschi86}, is normally given by
\begin{equation}\label{eq:totalp}
\mathcal{P}^{\nu} =  - \frac{1}{4 \pi} \int_S {\hat l}^{\nu} 
(\Psi_2^0 + \sigma_0 \dot {\bar \sigma}_0) dS^2 ,
\end{equation}
where
dot means $\frac{\partial}{\partial \hat u}$; i.e. the partial derivative with respect
to the Bondi time $\hat u$, $\Psi_2^0$ is a component of the Weyl tensor in the 
GHP\cite{Geroch73} notation and $\sigma_0$ is the leading order behavior of the spin
coefficient $\sigma$ in terms of the asymptotic coordinate $\hat r$
and $dS^2$ is the surface element of the unit sphere.
So the set of intrinsic Bondi coordinates are $(\hat u,\theta,\phi)$ or
$(\hat u,\zeta,\bar\zeta)$; where $(\zeta,\bar\zeta)$ are complex stereographic coordinates
of the sphere; which are related to the standard coordinates by
$\zeta = e^{i \phi} \cot(\frac{\theta}{2})$.

The total momentum for the monopole particle can be calculated using the
charge integral of the Riemann tensor technique.
In this subsection we will use the notation of reference \cite{Moreschi04}.

The total momentum and flux can be calculated in terms of charge integrals;
as described in \cite{Gallo:2016hpy,Gallo:2017yys}.
From which it can derive that
the time variation of the Bondi momentum, {of a generic spacetime,} 
is expressed by
\begin{equation}\label{eq:bondibalance}
\dot{\mathcal{P}}^{\mu} =  - \frac{1}{4 \pi } \int_S 
{\hat l}^{\mu} \sigma'_0 {\bar \sigma}'_0 dS^2 
\equiv -\mathcal{F}^{\mu}
;
\end{equation}
that is, $\mathcal{F}^{\mu}$ is the total momentum flux.

In reference \cite{Gallo:2016hpy} we have discussed how to describe
the total momentum and flux in terms of a general time coordinate.
Let $u=$constant represent a general time coordinate and set of sections
such that
\begin{equation}
 \frac{\partial \tilde u}{\partial u} 
= \tilde V(\tilde u,\zeta,\bar\zeta)
= \tilde V(u,\zeta,\bar\zeta) > 0 ,
\end{equation}
is the time derivative of the inertial (Bondi) time $\tilde u$ 
with respect to the non-inertial time $u$;
so that we can use the Bondi coordinates or the non-inertial coordinates
$(u,\zeta,\bar\zeta)$.
Then, following \cite{Gallo:2016hpy}, the general flux expression is
\begin{equation}\label{eq:bondibalanceV}
\frac{d\mathcal{P}^{\mu} }{du}
 =  - \frac{1}{4 \pi } \int_S 
{\hat l}^{\mu} \tilde V \, \sigma'_0 \, {\bar \sigma}'_0 \, dS^2 
= -\mathcal{F}_V^{\mu}
;
\end{equation}
where now $\mathcal{F}_V^{\mu}$ is the instantaneous momentum flux
with respect to the time $u$.

\section{Monopole geometry from asymptotic structure}\label{sec:mono}
\subsection{The metric}
If we take the general asymptotic form of an adapted null tetrad
of an asymptotically flat spacetime\cite{Moreschi87} and keep only the terms
associated with a monopole, then one is left with the line element
\begin{equation}\label{eq:monopole-linelement}
ds^2 = \left( - 2 \frac{\dot{V}}{V} r  + K_{V} - 2\frac{M(u)}{r}\right) du^{2} 
 + 2 \; du \;dr-  \frac{r^{2}}{P^2} d\zeta \;d\bar{\zeta} , 
\end{equation}
where $P=P(u,\zeta ,\bar{\zeta})=V\, P_0$, 
and $K_{V}$ is defined in equation (\ref{eq:kv}).
We refer to this as the monopole particle line element and is our model for
the sub-metric (\ref{eq:gA}).

This line element of a monopole can be related to 
what we have called Robinson-Trautman(RT) geometries\cite{Dain96}; which are generalizations
of Robinson-Trautman spacetimes.
Robinson-Trautman\cite{Robinson62} spacetimes have been very useful 
for estimating the total gravitational radiation in the head-on black hole
collision\cite{Moreschi96}\cite{Moreschi99}\cite{Anninos93}. 
In reference \cite{Moreschi96} we have applied these geometries to the 
description of
the total energy radiated in the head-on black hole collision with equal
mass; and it was shown that our calculations agree remarkably well with
the numerical exact calculations of Anninos et.al.\cite{Anninos93}.
The case of unequal mass black hole collision, was treated numerically
in reference \cite{Anninos98}; and our technique based on the use
of the RT geometries\cite{Moreschi99} showed again an impressive 
agreement with the exact calculations.

We assume that the mass parameter $M=\tt constant$
but 
to account for a possible time variation of the mass, we also
	consider the degree of freedom $\mu$;
so that instead of $M(u)$ we write $M \, \mu(u)$.

Then, the components $\xi^{i}$, $U$ and $X^{i}$ are then:
\begin{equation}
\xi ^{0}=0,\quad \xi ^{2}=\frac{\xi _{0}^{2}}{r},\quad \xi ^{3}=\frac{\xi
_{0}^{3}}{r}, 
\end{equation}
with
\begin{equation}\label{eq:xileadingRT}
\xi _{0}^{2}=\sqrt{2}P_{0}\;V,\qquad \xi _{0}^{3}=-i\xi _{0}^{2}; 
\end{equation}
\begin{equation}
U=rU_{00}+U_{0}+\frac{U_{1}}{r}, 
\end{equation}
where
\begin{equation}\label{eq:doblev}
U_{00}=\frac{\dot{V}}{V},\quad U_{0}=-\frac{1}{2}K_{V},\quad U_{1}= M \mu
, 
\end{equation}
where the curvature $K_V$ of the 2-metric appearing in equation 
(\ref{eq:desphere}), is given by (\ref{eq:kv}), $M$ is a constant
and
\begin{equation}
X^{0}=1,\quad X^{2}=0,\quad X^{3}=0;
\end{equation}
and where $\eth_V$ is the edth operator, in the GHP notation, 
of the sphere with metric 
(\ref{eq:desphere}).

There is only one Ricci spinor component different from zero, namely
\begin{equation}\label{eq:phi22}
 \Phi_{22} = \frac{3 M \mu \frac{\dot V}{V} - M \dot \mu + \frac{1}{2} \overline{\eth}_V \eth_V K_V }{r^2} .
\end{equation}
Note the change of sign convention with respect to \cite{Dain96}.

It is important to remark here that for a 
general compact object one 
would have for $\Phi_{22}$ an asymptotic behavior of the form 
$\Phi_{22}= \frac{\Phi_{22}^0}{r^2} + \delta \Phi_{22}(u,r,\zeta,\bar\zeta)$ 
where $ \delta \Phi_{22}$ decays to zero faster than $\frac{1}{r^2}$ for large $r$.
Therefore, in general, a condition on $\Phi_{22}^0$ would not imply
direct restrictions in the interior of the spacetime.
However, equation (\ref{eq:phi22}), tells us that in this monopole geometry
a condition on $\Phi_{22}^0$, which in general would be an asymptotic
condition,
 is a direct restrictions on the motion
of the central object, since there is only one single $r$ term dependence.
This is a very nice and peculiar property of this gauge.

In this presentation we are going to assume a constant mass particle
model; so that from now on we take $\mu=1$.

\subsection{Global quantities of this geometry}
Let us recall that the total momentum for RT geometries
can be expressed\cite{Dain96} as
\begin{equation}
 \mathcal{P}^\mu = \frac{1}{4 \pi} 
\int 
\, \frac{ M }{V^3} \hat l^\mu dS^2 
.
\end{equation}

It is also very interesting to calculate\cite{Dain96} the time variation of the total
momentum in these geometries. 
With respect to the instantaneous inertial (Bondi) time $\tilde u$ one has
\begin{equation}\label{eq:dPduB}
 \frac{d\mathcal{P}^\mu}{d\tilde u} = -\frac{1}{4 \pi} \int 
\left(
\frac{\partial \sigma_0}{\partial \tilde u} \frac{\partial \bar\sigma_0}{\partial \tilde u}
 + \Phi_{(B)22}^0
\right) \hat l^\mu dS^2 ;
\end{equation}
while the time derivative of the total momentum with respect to the RT
time is
\begin{equation}\label{eq:dPdu}
 \frac{d\mathcal{P}^\mu}{du} = -\frac{1}{4 \pi} \int 
\left(
\frac{\eth^2 V \bar\eth^2 V}{V}  + \frac{\Phi_{22}^{0}}{V^3}
\right) \hat l^\mu dS^2 .
\end{equation}
It is also convenient to recall the relations between the inertial  quantities
and the intrinsic ones, namely:
\begin{equation}
 \frac{\partial \sigma_0}{\partial \tilde u} = \frac{\eth^2 V}{V} 
= - \bar\sigma'_0
,
\end{equation}
\begin{equation}
 \Phi_{(B)22}^0 = \frac{\Phi_{22}^{0}}{V^4} ,
\end{equation}
and
\begin{equation}\label{eq:duBdu}
 \frac{\partial \tilde u}{\partial u} = V .
\end{equation}
Note that what it was called $\tilde{V}$ in the discussion of general
asymptotically flat spacetimes, of section \ref{sec:asympt}, becomes
in this geometry just $V$.

When viewing monopole particles as RT geometries the time parameter
$u$ coincides with the proper time of the particle.

Using the proper time $u$, one sees that demanding that $\frac{\Phi_{22}^{0}}{V^3}$
has no $l=0$ or $l=1$ spherical harmonic components, provides the appropriate
asymptotic
balanced equations of motion; namely
\begin{equation}\label{eq:dPdubalanced}
 \frac{d\mathcal{P}^\mu}{du} = -\frac{1}{4 \pi} \int 
\frac{\eth^2 V \bar\eth^2 V}{V}  
\hat l^\mu dS^2 ;
\end{equation}
we discuss this in detail below.

In a sense,
the last equation is actually an illusion since the Bianchi identities are really
identities. The prefer to take the 
non-trivial equation as that imposed on $\Phi_{22}$.

\section{Monopole dynamics in the null gauge}\label{sec:dynamics}

\subsection{Dynamics from the balanced equations of motion approach}

We have discussed elsewhere\cite{Gallo:2016hpy} how to construct balanced equations of motion
in a general setting applicable to different gauges.
Having the flux of momentum as described by the asymptotic balanced equations
as given by (\ref{eq:dPdubalanced}),
we set the 
flux force by
\begin{equation}\label{eq:flujo}
\begin{split}
 F(u')^\mu =& - \Upsilon \mathcal{F}_V^{\mu}
=
-\frac{\Upsilon}{4 \pi} \int \frac{\eth^2 V \bar\eth^2 V}{V}  \hat l^\mu dS^2 \\
 =&
 \frac{\Upsilon}{4 \pi} \int \frac{1}{2V^3} \bar\eth_V \eth_V K_V  \hat l^\mu dS^2
;
\end{split}
\end{equation}
where we are denoting with $u'$ the asymptotic time related to the
proper time $\tau$, as explained in \cite{Gallo:2016hpy}.
The second equality comes from the following property:
\begin{equation}\label{eq:propiedadKV}
\begin{split}
\frac{1}{2V^3} \bar\eth_V \eth_V K_V  
=&
\frac{1}{2 V^3} V^2 \bar\eth \eth K_V  \\
=&
\frac{1}{2 V^3}
\big(
2 V^3 \bar\eth^2 \eth^2 V -  2 V^2 \bar\eth^2 V \eth^2 V
\big) \\
=&
\,  \bar\eth^2 \eth^2 V -   \frac{1}{ V} \bar\eth^2 V \eth^2 V
;
\end{split}
\end{equation}
so that the first term on the right hand side does not have $l=0$
and $l=1$ contributions, in an expansion with spherical harmonics.

There are two natural dynamical times in the interior, for particle $A$;
one is $\tau$, the proper time with respect to the metric $g_B$,
and the other is the proper time $\tau_0$,
with respect to the metric $\eta$.
Let us denote with  $\mathbf{v}$ and $v$ be the corresponding tangent vectors
to the proper times $\tau_0$ and $\tau$ respectively.
Then the basic differential operators are $\mathbf{v}^b \partial_b \mathbf{v}^a$
or ${v}^b \nabla_{(B)b} {v}^a$;
where $\partial_b$ is the covariant derivative associated with the metric $\eta$.

Let us note that the two velocity vectors are proportional
\begin{equation}
v^b = \Upsilon \mathbf{v}^b .
\end{equation}
Notice that
\begin{equation}
\frac{d\tau}{d \tau_0}= \frac{1}{\Upsilon}  .
\end{equation}
We can use
\begin{equation}
1 = g_B(v,v) = \Upsilon^2 g_B(\mathbf{v},\mathbf{v})
= \Upsilon^2 \big( 1 +  h_B(\mathbf{v},\mathbf{v}) \big)
;
\end{equation}
which gives $\Upsilon$ in terms of $\mathbf{v}$ and $g_B$.
Note that one expects $1/\Upsilon= \sqrt{1 +  h_B(\mathbf{v},\mathbf{v})} < 1$.

Expressing the covariant derivative $\nabla_{(B)a}$ of $g_B$, in terms of the covariant 
derivative $\partial_a$ of $\eta$, one has
\begin{equation}
\nabla_{(B)a} \, v^b = \partial_a \, v^b + \gamma^{\;b}_{a\;\;c} \,  v^c 
;
\end{equation}
and using the relation between the vectors $v$ and $\mathbf{v}$ one also has
\begin{equation}
\begin{split}
v^a \nabla_{(B)a} \, v^b =& 
\Upsilon \mathbf{v}^a \nabla_{(B)a} \, \Upsilon  \mathbf{v}^b \\
=& 
\Upsilon \mathbf{v}^a \partial_a \, \Upsilon  \mathbf{v}^b
+ \Upsilon^2 \gamma^{\;b}_{a\;\;c} \, \mathbf{v}^a \,  \mathbf{v}^c \\
=& 
\Upsilon^2 \mathbf{v}^a \partial_a \, \mathbf{v}^b 
+ \Upsilon \frac{d\Upsilon}{d\tau_0} \, \mathbf{v}^b
+ \Upsilon^2 \gamma^{\;b}_{a\;\;c} \, \mathbf{v}^a \,  \mathbf{v}^c 
.
\end{split}
\end{equation}

Then, using the general treatment of the balanced approach\cite{Gallo:2016hpy}, 
one can write
 at the  time $u'$
 the equations of motion it terms of
this flux vector as
\begin{equation}\label{eq:balanced-flat4}
\begin{split}
M \Big(&
 \frac{1}{\big(\frac{du'}{d\tau_0}\big)^2} \mathbf{v}^a \partial_a \, \mathbf{v}^b 
+ \frac{1}{\big(\frac{du'}{d\tau_0}\big)^2} \gamma^{\;b}_{a\;\;c} \, \mathbf{v}^a 
               \,  \mathbf{v}^c \\
& - \frac{1}{\big(\frac{du'}{d\tau_0}\big)^3} \frac{d^2 u'}{d\tau_0^2} \mathbf{v}^b
 + w \frac{1}{\big(\frac{du'}{d\tau_0}\big)} \mathbf{v}^b  \Big)(\tau_0) 
= {F}^b(u', v) 
; 
\end{split}
\end{equation}
where we note an interior degree of freedom $w$.
The null gauge approach provides with a simple relation between
the inner dynamical time $\tau$ with the asymptotic dynamical time
$u'$; since one has $u'=\tau$.
In other words, in this first version of the model in the null gauge,
we neglect a possible degree of freedom, discussed in \cite{Gallo:2016hpy}
by taking $\frac{du'}{d\tau}=1$; so that $\frac{du'}{d\tau_0}=1/\Upsilon$.
In this work we reserve to denote with $u$ the asymptotic null coordinate
that is determined by $\tau_0$, at the world line.
This in turn allows us to express the equations of motion as
\begin{equation}\label{eq:balanced-flat-3}
\begin{split}
M \Big(&
  \mathbf{v}^a \partial_a \, \mathbf{v}^b 
+  \gamma^{\;b}_{a\;\;c} \, \mathbf{v}^a  \,  \mathbf{v}^c \\
& 
+ \frac{1}{\Upsilon}  \frac{d\Upsilon}{d\tau_0} \, \mathbf{v}^b
 + \frac{w}{\Upsilon}  \mathbf{v}^b  \Big)(\tau_0) 
=  \frac{1}{\Upsilon} \mathbf{F}_0^\mu 
;
\end{split}
\end{equation}
where the momentum flux is calculated in terms of the proper time $\tau_0$.

Thus, equation (\ref{eq:balanced-flat-3}) are the main equations of motion.
It is natural to decompose this equation in the direction of $\mathbf{v}^d$
and its orthogonal complement.
Then, contracting this equation with $\eta_{bd} \mathbf{v}^d$ gives
\begin{equation}\label{eq:balance-con-v}
\begin{split}
M \Big(
\gamma^{\;b}_{a\;\;c} \, \mathbf{v}^a 
\,  \mathbf{v}^c \, \eta_{bd} \mathbf{v}^d 
+ 
\frac{1}{\Upsilon}  \frac{d\Upsilon}{d\tau_0}
&+  \frac{w}{\Upsilon}   
\Big)
=
\frac{1}{\Upsilon}  
\mathbf{F}_0^b \eta_{bd} \mathbf{v}^d
;
\end{split}
\end{equation}
and it remains the equations of motion
\begin{equation}\label{eq:balanced-flat3}
\begin{split}
M 
\mathbf{a}^b  
= M  \mathbf{f}_\perp^b  
+ 
\frac{1}{\Upsilon}  
\mathbf{F}_0^d
\big(
\eta_d^{\;\;b} - \mathbf{v}_d \mathbf{v}^b
\big)
;
\end{split}
\end{equation}
which is  orthogonal to $\mathbf{v}^a$,
where
\begin{equation}
\mathbf{a}^a \equiv \mathbf{v}^b \partial_b \mathbf{v}^a ,
\end{equation}
and $\mathbf{f}_\perp^b$ is defined by
\begin{equation}
\mathbf{f}_\perp^b \equiv - \gamma^{\;d}_{a\;\;c} \, \mathbf{v}^a \,  \mathbf{v}^c 
\big(
\eta_d^{\;\;b} - \mathbf{v}_d \mathbf{v}^b
\big)
;
\end{equation}
which, it should be remarked, only depends on the background 
geometry $g_B$ and $\mathbf{v}$;
and let us recall that, in general, the radiation force is given by
\begin{equation}\label{eq:force_tau_0}
 \mathbf{F}_0^\mu 
 = - \frac{1}{4 \pi } \int_S 
{\hat l}^{\mu} \ V \, \sigma'_0 \, {\bar \sigma}'_0 \, dS^2 
;
\end{equation}
with
\begin{equation}\label{eq:vee}
 V =  \frac{\partial \tilde u}{\partial \tau_0} 
;
\end{equation}
since we are using (\ref{eq:duBdu}) with $u=\tau_0$.

Equation (\ref{eq:balance-con-v}) is, in the general framework, 
understood as an equation for 
$w$.

We introduce here a notation that will be useful in the following discussions.
Let us define
\begin{equation}
\begin{split}
\mathbf{f}^b 
\equiv
-  \gamma^{\;b}_{a\;\;c} \, \mathbf{v}^a \,  \mathbf{v}^c 
- \frac{1}{\Upsilon}  \frac{d\Upsilon}{d\tau_0} \, \mathbf{v}^b
- \frac{w}{\Upsilon}  \mathbf{v}^b
;
\end{split}
\end{equation}
so that the equations of motion can be expressed as
\begin{equation}\label{eq:mot-compact}
\mathbf{a}^a =  \mathbf{f}^a + f_\lambda^a .
\end{equation}
with $ f_\lambda^\mu$ defined by
\begin{equation}\label{eq:flambda-F0}
M f_\lambda^\mu \equiv \frac{1}{\Upsilon} \mathbf{F}_0^\mu .
\end{equation}	

We will also use the notation 
\begin{equation}\label{eq:acel-scalar}
\mathcal{A} \equiv \mathbf{a}^{\mu} {l}_{\mu} ,
\end{equation}
and
\begin{equation}\label{eq:acel-geodesic-scalar}
\mathcal{F} \equiv \mathbf{f}^{\mu} {l}_{\mu} .
\end{equation}
Then, defining the scalar
\begin{equation}
f_\lambda = f_\lambda^\mu {l}_{\mu} = \frac{1}{M \Upsilon} \mathbf{F}_0^\mu {l}_{\mu} ;
\end{equation}
one can write the equations of motion in scalar form by
\begin{equation}\label{eq:AigFmasflambda}
\mathcal{A} = \mathcal{F} + f_\lambda .
\end{equation}
	
Note that due to (\ref{eq:balanced-flat3}) we can also express the
equations of motion by
\begin{equation}\label{eq:AigFmasflambda-perp}
\mathcal{A} = \mathcal{F}_\perp + f_{\lambda\perp} ;
\end{equation}
where
\begin{equation}
\mathcal{F}_\perp = \mathbf{f}_\perp^\mu \, l_\mu ,
\end{equation}
and
\begin{equation}
f_{\lambda\perp} = 
\frac{1}{M \Upsilon}  
\mathbf{F}_0^\nu
\big(
\eta_\nu^{\;\;\mu} - \mathbf{v}_\nu \mathbf{v}^\mu 
\big)  \, l_\mu
.
\end{equation}

With respect to the scalar of the geometry one can also express
\begin{equation}\label{eq:force_tau_0-1}
\begin{split}
\mathbf{F}_0^\mu 
=
 -\frac{1}{4 \pi} \int \frac{\eth^2 V \bar\eth^2 V}{V}  \hat l^\mu dS^2 ;
\end{split}
\end{equation}
and due to (\ref{eq:flujo}) we can also express it as
\begin{equation}\label{eq:force_tau_0-2}
\begin{split}
\mathbf{F}_0^\mu 
=
 \frac{1}{4 \pi} \int \frac{1}{2V^3} \bar\eth_V \eth_V K_V  \hat l^\mu dS^2
 .
\end{split}
\end{equation}

\subsection{The gravitational radiation degrees of freedom in the scalar $V$}

The discussion of the model uses different orders of different terms
appearing in the equations describing the geometry and governing
the dynamics.
For this purpose, let
us use the parameter $\gamma$ to denote the order of the gravitational constant,
and $\hat\gamma$ for some monotonic function of $\gamma$, to be determined
later.
The parameter $\hat\gamma$ can be thought as a measure of the effects
of gravitational radiation.

We decompose the scalar $V$ in terms of
\begin{equation}\label{eq:Vdecomposed}
 V = V_\eta (1 + \hat\gamma V_{\hat\gamma}) ;
\end{equation}
where $V_\eta$ is the conformal factor of the angular part with respect to the flat
background frame, and $V_{\hat\gamma}$ has information of first order in
gravitational radiation.

In the construction of this model there arises more than one natural
dynamical time; from which at least three of them are essential.
The two proper times associated to the  metrics $\eta$ and $g_B$, in the interior,
and the null global coordinate time of the geometry determined $g_A$.

The coordinate time $u$ has been chosen so that, with respect to the 
geometry of $\eta$, $u=\tau_0$ at $r=0$. That is, $\tau_0$ labels the null
hypersurfaces $u={\tt constant}$.

\subsection{Relation with the field equation of the monopole model}
A peculiarity of the null gauge is that the field equations provide with a 
straightforward direct connection between the interior,
where the equations of motions are used,
 and the asymptotic region, where the flux of momentum is calculated.
Therefore, beyond the general approach for balanced dynamics that
we have been using; in this case one has available more relations
that can be used to improve in the construction of the 
balanced equations of motion.

One can see from equation (\ref{eq:dPdu}), that to obtain the
asymptotic balance equation, one must demand that the $l=0$ and $l=1$ terms of
$\frac{\Phi_{22}^{0}}{V^3}$ must vanish; in other words
one must satisfy equation
\begin{equation}\label{eq:dPdu-Phi}
\int  \frac{\Phi_{22}^{0}}{V^3} \;  \hat l^\mu \, dS^2 = 0 
;
\end{equation}
which is our main equation from the asymptotic structure.

Reading the leading order behavior $\Phi_{22}^0$ from (\ref{eq:phi22});
we can express the complete monopole field equation as:
\begin{equation}\label{eq:vacuum1}
 3 M  \frac{\dot V}{V^4} 
= -\frac{1}{2V^3} \bar\eth_V \eth_V K_V  
.
\end{equation}
Let us remark that the $l=0$ and $l=1$ terms of the left hand side 
of (\ref{eq:vacuum1}) constitute {\sf minus} 
the time derivative of the total momentum.
Then, taking into account property (\ref{eq:propiedadKV}), we conclude that
it is equivalent
to solve equation (\ref{eq:dPdu-Phi}) than the asymptotic balance
equation (\ref{eq:dPdubalanced}).
In other words, the $l=0$ and $l=1$ terms of the right hand side of
(\ref{eq:vacuum1}) is precisely the radiation flux of total momentum;
as corroborated by equation (\ref{eq:force_tau_0-2}).
So that in principle we can use the higher order terms of the left hand side
of (\ref{eq:vacuum1}) to improve on the left hand side of
the equations of motion (\ref{eq:balanced-flat4});
this is what is done below.
This is another subtle advantage of the use of the null gauge.

\subsection{Properties of the monopole in the null gauge}
Let us summarize here what is our approach for the monopole and then
remark some convenient properties.

We have taken our definition of the monopole from the leading order behavior
of a general asymptotically flat spacetime\cite{Moreschi87}.
The Ricci tensor has only one null component different from zero;
namely equation (\ref{eq:phi22}).
The asymptotic balance equation, for the total momentum is the requirement
of the $l=0$ and $l=1$ terms of equation (\ref{eq:vacuum1});
where it should be remarked that (\ref{eq:vacuum1}) is proportional to (\ref{eq:phi22}).

Regarding the properties of this representation of the monopole,
let us start by mentioning that
in the decomposition (\ref{eq:Vdecomposed}), the conformal factor $V_\eta$
has the information of the motion of the monopole with respect to the chosen
flat background, that in our case is the center of mass frame.
Then in general, the time evolution of $V_\eta$ will have the information
of an accelerated motion of our body $A$ due to the existence of the
rest of the system that we call $B$. What is the situation if
there is no system $B$? Then, the interior center of mass should
be identified with the asymptotic center of mass, defined in terms
of the total angular momentum\cite{Moreschi04};
which in terms of the null gauge, it means that, in the rest frame,
$V=1$+radiation terms.
Also, in the case there is no system $B$, the line element (\ref{eq:monopole-linelement})
represents the metric of an isolated body;
and the complete equation (\ref{eq:vacuum1}) is the Robinson-Trautman
equation, that describes a perturbed black hole.
We have studied in \cite{Frittelli92} how these spacetimes decay exponentially
fast to the Schwarzschild solution.
So, a very convenient property of this monopole description in the null gauge,
is that it represents a black hole that satisfies the exact
field equations, when isolated.

When it is not isolated, then the nature of equation (\ref{eq:vacuum1})
changes completely, as we will see below, since now the time evolution of $V_\eta$
becomes the driving term of the equation, and we now have an equation
for $V_{\hat{\gamma}}$ as expressed in (\ref{eq:Vdecomposed}).

The field equation can be expressed as
\begin{equation}\label{eq:vacuum1-b}
3 M  \frac{\dot V}{V^4} 
= -\frac{1}{2V^3} \bar\eth_V \eth_V K_V  
= -\,  \bar\eth^2 \eth^2 V +   \frac{1}{ V} \bar\eth^2 V \eth^2 V
;
\end{equation}
then, in the rest frame, one can replace
\begin{equation}\label{eq:vacuum1-c}
\begin{split}
3 M\frac{\dot V}{V^4} 
=&
3 M
\frac{{\dot{( V_\eta)}}   }{ 
	( 1 + \hat\gamma V_{\hat\gamma})^3} 
+ 
3 M
\frac{ \hat\gamma \dot V_{\hat\gamma} }
{ 
	( 1 + \hat\gamma V_{\hat\gamma})^4} \\
=& - \hat\gamma \,  \bar\eth^2 \eth^2 V_{\hat\gamma} 
+   \frac{\hat\gamma^2}{ ( 1 + \hat\gamma V_{\hat\gamma})} \bar\eth^2 V_{\hat\gamma} \eth^2 V_{\hat\gamma}
.
\end{split}
\end{equation}
We can in principle solve the coupled system of the balanced equation of motion
and the exact equation (\ref{eq:vacuum1-b}) numerically;
where given $V_{\hat{\gamma}}(u_0,\zeta, \bar{\zeta} )$ at an initial
time $u=u_0$ and $V_\eta(u_0 ,\zeta, \bar{\zeta} )$,
then one can integrate in the time domain.
Equation (\ref{eq:vacuum1-b}) is still a parabolic equation with a source
for $V_{\hat{\gamma}}$; so that irrespective of the choice for 
$V_{\hat{\gamma}}(u_0,\zeta, \bar{\zeta} )$, one expects that all solutions
will converge exponentially fast to a solution completely 
driven by the source acceleration $\dot V_\eta(u ,\zeta, \bar{\zeta} )$.
In a situation in which the radiation effects are bounded, one can
study the regime in which $\hat{\gamma} V_{\hat{\gamma}} < 1$, and
so deal with an expansion of the denominators in (\ref{eq:vacuum1-c});
so that one has
\begin{equation}\label{eq:vacuum1-c-b}
\begin{split}
3 M&
\dot V_\eta 
(1 - 3\hat\gamma V_{\hat\gamma} + 6 (\hat\gamma V_{\hat\gamma})^2
- 10 (\hat\gamma V_{\hat\gamma})^3 + \ldots ) \\
&+3 M \hat\gamma 
\dot V_{\hat\gamma}  
( 1 - 4 \hat\gamma V_{\hat\gamma} + 10 (\hat\gamma V_{\hat\gamma})^2
+ \ldots ) \\
=& - \hat\gamma\,  \bar\eth^2 \eth^2 V_{\hat\gamma} 
+   
\hat\gamma^2
\big(
1 - \hat\gamma V_{\hat\gamma} 
+ \ldots
\big)
\bar\eth^2 V_{\hat\gamma} \eth^2 V_{\hat\gamma}
.
\end{split}
\end{equation}
In this way we will try an approximate solution, that is driven
by the source acceleration.
The guiding idea is that the $l=0$ term of the right hand side is at lest
of order $\mathscr{O}(G^2)$, while 
the $l=1$ term is at lest of order $\mathscr{O}(G^3)$;
therefore, the left hand side is calculated accordingly.

\subsection{Decomposition of the scalars by their angular behaviour
	and time derivatives}

Given a function $H(u,\zeta,\bar{\zeta})$ on the future null cones,
one has the natural action of the Lorentz group on the angular 
coordinates\cite{Goldberg67,Held70}, which allows us make the decomposition
\begin{equation}\label{eq:autofunciones}
\bar\eth_{V_\eta} \eth_{V_\eta} H_l = - \frac{l(l+1)}{2} H_l ;
\end{equation}
for $l=0,1,2,...$; in terms of the edth operators of the
instantaneous rest frame.
With this, it is natural then to define the projection operators
$\mathcal{T}_l$ which have the property
$\mathcal{T}_l\big(H\big)= H_l$.

We use the decomposition for $V$ in terms of eigenfunctions of the operator 
$\bar\eth_{V_\eta} \eth_{V_\eta}$, that is:
\begin{equation}\label{eq:vdeveta-0}
 V = V_\eta \bigg( 1 + \hat\gamma V_{\hat\gamma} \bigg)
= V_\eta \bigg( 1 + \hat\gamma ( V_0 + V_1 + V_2 + V_3 + \ldots ) \bigg) 
;
\end{equation}
where the subindex $l$ denotes the angular behavior.

In particular, let us note that
\begin{equation}
 \bar\eth_V \eth_V \mathcal{A} 
=
\bigg( 1 + \hat\gamma V_{\hat\gamma} \bigg)^2
 \bar\eth_{V_\eta} \eth_{V_\eta} \mathcal{A} 
=
 - \bigg( 1 + \hat\gamma V_{\hat\gamma} \bigg)^2 \mathcal{A} ;
\end{equation}
that is, we are using:
\begin{equation}
 \bar\eth_{V_\eta} \eth_{V_\eta} \mathcal{A} 
=
 -  \mathcal{A} .
\end{equation}

In order to study the vacuum equation, it is convenient to make an analysis
in terms of the angular behaviour and also in terms of the first few orders
of the equation.

In the rest reference frame, one has that the time derivative of $V$ is
\begin{equation}
\begin{split}
 \dot V =& \dot V_\eta \bigg( 1 + \hat\gamma (V_0 +  V_1 + V_2 + V_3 + \ldots ) \bigg) \\
&+   \hat\gamma (\dot V_0 + \dot V_1 + \dot V_2 + \dot V_3 + \ldots )
.
\end{split}
\end{equation}
And also we will use the formal decomposition of the following expressions
\begin{equation}
\begin{split}
 \frac{1}{V^4} =& \frac{1}{ 
 	( 1 + \hat\gamma V_{\hat\gamma})^4 } \\
=& 
\big(  1 - 4\hat\gamma V_{\hat\gamma} + 10 (\hat\gamma V_{\hat\gamma})^2
    - 20 (\hat\gamma V_{\hat\gamma})^3 + \ldots \big)
,
\end{split}
\end{equation}
\begin{equation}
\begin{split}
 \frac{1}{V^3} =& \frac{1}{ 
 	( 1 + \hat\gamma V_{\hat\gamma})^3 } \\
=&  
\big(
1 - 3\hat\gamma V_{\hat\gamma} + 6 (\hat\gamma V_{\hat\gamma})^2
    - 10 (\hat\gamma V_{\hat\gamma})^3  
+ \ldots
    \big)
,
\end{split}
\end{equation}
and
\begin{equation}\label{ec:1sobreV}
\begin{split}
 \frac{1}{V} =& \frac{1}{
 	( 1 + \hat\gamma V_{\hat\gamma}) } \\
=& 
\big(
1 - \hat\gamma V_{\hat\gamma} +  (\hat\gamma V_{\hat\gamma})^2
    -  (\hat\gamma V_{\hat\gamma})^3 + \ldots
    \big)
.
\end{split}
\end{equation}
Then from the left hand side of (\ref{eq:vacuum1})  we have:
\begin{equation}
\begin{split}
 \frac{\dot V}{V^4} 
=&
\frac{{\dot{( V_\eta)}}   }{ 
	( 1 + \hat\gamma V_{\hat\gamma})^3} 
+ 
\frac{ \hat\gamma (\dot V_0 + \dot V_1 + \dot V_2 + \dot V_3 + \ldots ) }
{ 
	( 1 + \hat\gamma V_{\hat\gamma})^4} \\
=&
\dot V_\eta 
(1 - 3\hat\gamma V_{\hat\gamma} + 6 (\hat\gamma V_{\hat\gamma})^2
    - 10 (\hat\gamma V_{\hat\gamma})^3 + \ldots ) \\
&+ \hat\gamma 
(\dot V_0 + \dot V_1 + \dot V_2 + \dot V_3 + \ldots ) \\
& \; ( 1 - 4 \hat\gamma V_{\hat\gamma} + 10 (\hat\gamma V_{\hat\gamma})^2
    - 20 (\hat\gamma V_{\hat\gamma})^3 + \ldots )
\end{split}
\end{equation}
And from the right hand side of (\ref{eq:vacuum1})  we have:
\begin{equation}
\begin{split}
\frac{1}{V^3} \bar\eth_V & \eth_V K_V
=
\frac{1}{ 
	( 1 + \hat\gamma V_{\hat\gamma})}
 \bar\eth_{V_\eta} \eth_{V_\eta} K_V \\
=&
\frac{1}{ 
	( 1 + \hat\gamma V_{\hat\gamma})}
\big( - K_{V1} - 3 K_{V2} - 6 K_{V3}  + \ldots \big) \\
=&
\big( - K_{V1} - 3 K_{V2} - 6 K_{V3}  + \ldots \big) \\
& \; ( 1 -  \hat\gamma V_{\hat\gamma} +  (\hat\gamma V_{\hat\gamma})^2
    -  (\hat\gamma V_{\hat\gamma})^3 + \ldots )
;
\end{split}
\end{equation}
where we are also using the decomposition of $K_V$ in terms of the 
eigenfunctions of the operator $\bar\eth_{V_\eta} \eth_{V_\eta}$.

Let us recall that from (\ref{eq:kv}), that one can also express
\begin{equation}\label{eq:kveta}
K_{V}
=
 \frac{2 
 	\big( 1 + \hat\gamma V_{\hat\gamma} \big)}{V_\eta}
 ~\overline{\eth }_{V_\eta}\eth _{V_\eta}\, V
-\frac{2}{V_\eta^{2}}~\eth _{V_\eta} V~%
\overline{\eth }_{V_\eta}V + V^{2} ,
\end{equation}
showing the explicit dependence of the edth operators on $V_\eta$.

\subsection{The first angular terms of the field equation}

In this construction the acceleration $\mathcal{A}$, 
produced by the `force' $\mathcal{F}$, due to the presence
of other systems, provokes the emission of gravitational radiation.
In other words, in equation (\ref{eq:vacuum1}) there would be no radiation
and therefore no back reaction if the were not external force $\mathcal{F}$;
so that it should be consider as the source of the radiation effects.

We study next the way to connect the external force to the radiation
degrees of freedom.

In this first version of the null gauge model, we assume that it is only
necessary to consider up to the quadrupole structure of the spacetime;
in other words,
we assume we can build the geometry in terms of $V_0$, $V_1$ 
and $V_2$, assuming that all other $V_l$'s with higher $l$ are negligible.
Then expanding (\ref{eq:vacuum1}), in the instantaneous rest frame, in the first few orders one finds
\begin{equation}\label{eq:vacuum2}
\begin{split}
3 M \frac{\dot V}{V^4} 
=& 3 M 
\bigg(
\dot V_\eta 
+
\hat\gamma \big[
- 3 \dot V_\eta (V_0 + V_1 + V_2) \\
& \qquad \qquad \qquad +
(\dot V_0 + \dot V_1 + \dot V_2 )
\big] \\
&+
\hat\gamma^2 \big[
6 \dot V_\eta (V_0 + V_1 + V_2)^2 \\
& \;\;\; - 4(\dot V_0 + \dot V_1 + \dot V_2 )(V_0 + V_1 + V_2)
\big] + \ldots
\Bigg) \\
=&
\frac{1}{2}
\big(  K_{V1} + 3 K_{V2} + 6 K_{V3}  + \ldots \big) \\
& \; ( 1 -  \hat\gamma (V_0 +  V_1 + V_2 ) +  \hat\gamma^2 (V_0 +  V_1 + V_2 )^2
    + \ldots ) 
.
\end{split}
\end{equation}

In this setting, we will assume that time derivatives of $V_{\hat{\gamma}}$
increases the order of the term, so that considering terms up to the three order level,
the left hand side of expression (\ref{eq:vacuum2}) is
\begin{equation}\label{eq:vacuum2-lfh-con-V0-2}
\begin{split}
3 M \frac{\dot V}{V^4} 
=& 3 M 
\bigg(
\dot V_\eta 
+
\hat\gamma \big[
- 3 \dot V_\eta (V_0 + V_1 + V_2) \\
& \qquad \qquad \qquad +
(\dot V_0 + \dot V_1 + \dot V_2 )
\big] \\
&+
\hat\gamma^2 
6 \dot V_\eta (V_0 + V_1 + V_2)^2  + \mathscr{O}(\hat\gamma^3)
\bigg) 
.
\end{split}
\end{equation}

In order to calculate the terms that contribute to the $l=0$ and
$l=1$ spherical harmonic decomposition of the right hand side
of equation (\ref{eq:vacuum2}), we can either use the original
expression appearing in the field equation or we can use
\begin{equation}\label{eq:edth2edth2bsobrev}
\frac{\eth^2 V \bar\eth^2 V}{V} ;
\end{equation}
as appears in (\ref{eq:flujo}).
This expression looks simpler to handle since at the instantaneous
rest frame one has
\begin{equation}
\bar\eth^2 V = \hat \gamma \bar\eth^2 (V_2 + V_3 + \cdots) ;
\end{equation}
and since we are neglecting in this first null model higher order
angular behavior, we just have
\begin{equation}
\bar\eth^2 V = \hat \gamma \bar\eth^2 V_2  .
\end{equation}
The complication is that we have to deal with spin weighted quantities.
The other factor is
\begin{equation}\tag{\ref{ec:1sobreV}}
\begin{split}
\frac{1}{V} 
=& 
\big(
1 - \hat\gamma V_{\hat\gamma} +  (\hat\gamma V_{\hat\gamma})^2
-  (\hat\gamma V_{\hat\gamma})^3 + \ldots
\big)
.
\end{split}
\end{equation}
Then, one can see that the contribution to the $l=0$ term of (\ref{eq:edth2edth2bsobrev})
comes from
\begin{equation}\label{eq:edth2edth2}
\hat \gamma^2 \bar\eth^2 V_2  \eth^2 V_2  ;
\end{equation}
while the first order contribution to the $l=1$ term comes from
\begin{equation}\label{eq:v1edth2edth2}
- \hat \gamma^3 V_1 \bar\eth^2 V_2  \eth^2 V_2  .
\end{equation}

\subsection{Combining the balanced equations of motion with the monopole
	field equation}

In the above equations we have suggested that one can solve the monopole
field equation by computing numerically the evolution of $V_{\hat{\gamma}}$
with high precision.
However, it is reasonable to also present an expansion of the monopole
field equation in terms of orders, by the choice of a particular sub-gauge.
In what follows we present a way to carryout this type of expansion,
that we believe will be useful in future calculations.

When analyzing the field equation (\ref{eq:vacuum1}), 
we use a decomposition of the scalar $V$, 
given in (\ref{eq:Vdecomposed}), which has information
of the interior equations of motion, as given by (\ref{eq:AigFmasflambda})
which in turn uses the force expression, as given by (\ref{eq:force_tau_0-2}).
It is important at this stage to recall a property of expressions as (\ref{eq:force_tau_0-2})
noted in \cite{Moreschi98}; which says the following:
defining 
\begin{equation}
\hat{\mathbf{F}}_0 = \mathbf{F}_0^\mu \hat l_\mu = \mathbf{F}_0^0 - \hat{\mathbf{F}}_0^{(3)}
,
\end{equation}
with 
$\hat{\mathbf{F}}_0^{(3)} = \mathbf{F}_0^1 \, \hat l^1
+ \mathbf{F}_0^2 \, \hat l^2 + \mathbf{F}_0^3 \, \hat l^3$,
then one has that
\begin{equation}
\frac{1}{2V^3} \bar\eth_V \eth_V K_V = 
\mathbf{F}_0^0 - 3 \, \hat{\mathbf{F}}_0^{(3)} + \text{terms with }\hat Y_{l_2,m} ,
\end{equation}
with $l_2 \geqslant 2$, with respect to the angular variables of the inertial
system.
This is important to notice since we will use equation (\ref{eq:AigFmasflambda})
in the form
\begin{equation}\label{eq:AigFmasflambda-2}
\mathcal{A} = \mathcal{F} + 
\frac{1}{M \Upsilon V_\eta} \hat{\mathbf{F}}_0
.
\end{equation}
Note that at the instantaneous rest frame one has
\begin{equation}\label{eq:AigFmasflambda-2-0}
\mathcal{F}^0 = 
-
\frac{1}{M \Upsilon} \hat{\mathbf{F}}_0^0
.
\end{equation}

Replacing $\dot V_\eta \longrightarrow \mathcal{A}$ in (\ref{eq:vacuum2}),
in the instantaneous rest frame, one obtains
\begin{equation}
\begin{split}
& 3 M 
\bigg(
\big( \mathcal{F} + 
\frac{1}{M \Upsilon } (\mathbf{F}_0^0 - \hat{\mathbf{F}}_0^{(3)}) \big)
(1 - 3\hat\gamma V_{\hat\gamma} + 6 (\hat\gamma V_{\hat\gamma})^2
+ \ldots ) \\
&+ \hat\gamma (\dot V_0 + \dot V_1 + \dot V_2 ) \\
& \;( 1 - 4 \hat\gamma V_{\hat\gamma} + 10 (\hat\gamma V_{\hat\gamma})^2
- 20 (\hat\gamma V_{\hat\gamma})^3 + \ldots )
\Bigg) \\
=&
 3 M 
\bigg(
 - \hat{\mathcal{F}}^{(3)}  
(1 - 3\hat\gamma V_{\hat\gamma} + 6 (\hat\gamma V_{\hat\gamma})^2
+ \ldots ) \\
&+ \hat\gamma (\dot V_0 + \dot V_1 + \dot V_2  ) \\
& \;( 1 - 4 \hat\gamma V_{\hat\gamma} + 10 (\hat\gamma V_{\hat\gamma})^2
- 20 (\hat\gamma V_{\hat\gamma})^3 + \ldots )
\Bigg) \\
&-  \frac{3}{ \Upsilon }  \hat{\mathbf{F}}_0^{(3)} \\
=&
-\frac{1}{2}
\frac{1}{V^3} \bar\eth_V \eth_V K_V \\
=& -( \mathbf{F}_0^0 - 3 \, \hat{\mathbf{F}}_0^{(3)} ) + \text{terms with } \hat Y_{l_2,m} 
;
\end{split}
\end{equation}
where we have used that 
$\mathcal{F}= -\frac{1}{M\Upsilon} \hat{\mathbf{F}}_0^0 - \hat{\mathcal{F}}^{(3)}$
and we have neglected orders higher than the radiation terms.
Therefore, up to the third order one has
\begin{equation}\label{eq:vacuum2-fieldfh-con-V0}
\begin{split}
& 3 M 
\bigg(
(- \hat{\mathcal{F}}^{(3)})
+
\hat\gamma \big[
- 3 (- \hat{\mathcal{F}}^{(3)}) (V_0 + V_1 + V_2) \\
& \qquad \qquad \qquad +
(\dot V_0 + \dot V_1 + \dot V_2 )
\big] \\
&+
\hat\gamma^2 
6 (- \hat{\mathcal{F}}^{(3)}) (V_0 + V_1 + V_2)^2  + \mathscr{O}(\hat\gamma^3)
\bigg) \\
=&
 - \mathbf{F}_0^0 
 + 3 \, \hat{\mathbf{F}}_0^{(3)}  \big( 1 +  \frac{1}{ \Upsilon } \big)
 + \text{terms with }\hat Y_{l_2,m} 
;
\end{split}
\end{equation}

\subsection{Choosing an appropriate frame and ansatz}

We are going to use both, the notation in terms of vectors,
and also in terms of spherical harmonics.

Let us denote with $v_1^\mu$ the four vector such that
\begin{equation}
V_1 = v_1^\mu l_\mu = \frac{v_1^\mu \hat l_\mu}{V_\eta} ;
\end{equation}
then, since $V_1$ satisfies (\ref{eq:autofunciones}) with $l=1$,
we know that $v_1^\mu$ is orthogonal to $\mathbf{v}^\mu$.

Without loss of generality, we could assume that in the instantaneous
rest frame, the acceleration, at zero order, is given by
\begin{equation}
\mathcal{A} =  \mathsf{a} Y_{10} ;
\end{equation}
where, in order to understand the detail of the notation,
it should be remarked that if $\mathbf{a}^\mu$ were the four vector
that corresponds to an acceleration $a^z$ in the $z$ direction, then
one would have
\begin{equation}
\mathcal{A} = - a^z \sqrt{\frac{4 \pi}{3}} Y_{10} .
\end{equation}
Actually, the source of any acceleration is $\mathcal{F}$;
so that it is more convenient to assume that
\begin{equation}
\mathcal{F} = a Y_{10} + \mathcal{F}^0 ;
\end{equation}
that  is
\begin{equation}
- \hat{\mathcal{F}}^{(3)} = a Y_{10} .
\end{equation}

In order to proceed with the calculation it is needed to determine
in more detail the relation of the gravitational wave degree of
freedom with the mechanical ones.
We now take as an ansatz that in this frame, we only need to consider
\begin{equation}
V_0 = b_0 Y_{00} ,
\end{equation}
\begin{equation}
V_1 = b_1 Y_{10} ,
\end{equation}
and
\begin{equation}
V_2 = b_2 Y_{20} .
\end{equation}

With this we can advance in the calculation of 
(\ref{eq:vacuum2-fieldfh-con-V0})
to
obtain the first orders of the  left hand side of the equation.

Note that now we have
\begin{equation}\label{eq:vacuum2-fieldfh-con-V0-2}
\begin{split}
& 3 M 
\bigg(
a Y_{10}
+
\hat\gamma \big[
- 3 \, a Y_{10} (b_0  Y_{00} + b_1  Y_{10} + b_2  Y_{20}) \\
& \qquad \qquad \qquad +
(\dot V_0 + \dot V_1 + \dot V_2 )
\big] \\
&+
\hat\gamma^2 
6\, a Y_{10} (b_0  Y_{00} + b_1  Y_{10} + b_2  Y_{20})^2  + \mathscr{O}(\hat\gamma^3)
\bigg) \\
=&
- \mathbf{F}_0^0 
+ 3 \, \hat{\mathbf{F}}_0^{(3)}  \big( 1 +  \frac{1}{ \Upsilon } \big)
+ \text{terms with }\hat Y_{l_2,m} 
.
\end{split}
\end{equation}

Using the properties of multiplication of spin-weighted spherical harmonics,
recalled in appendix \ref{ap:sph-harmo}, we find that 
the $l=0$ term of the left hand side of (\ref{eq:vacuum2-fieldfh-con-V0-2}) is
\begin{equation}
\begin{split}\label{eq:lfh-l0-2}
3 M 
&
\hat\gamma \big(
- 3   \frac{a b_1}{\sqrt{4 \pi}} Y_{00}  
+ \dot V_0 + \mathscr{T}_0(\dot V_1 ) 
\big) 
+ M \mathscr{O}(\hat{\gamma}^2)
\end{split}
\end{equation}
while the $l=1$ term
of the left hand side of the equation is
\begin{equation}
\begin{split}\label{eq:lfh-l1-2}
3 M 
\Bigg[&
a Y_{10}
+
\hat\gamma \Big(
- 3 ( \frac{a b_0}{\sqrt{4 \pi}} + \frac{a b_2}{\sqrt{5 \pi}} ) Y_{10} 
+ \mathscr{T}_1( \dot V_1 )
\Big) \\
+&
\hat{\gamma}^2 6
\frac{1}{\pi} 
\bigg(
\frac{1}{4} 
a \,b_0^2  + \frac{1}{\sqrt{5}} a \, b_0 b_2    \\
&\qquad \quad + \frac{9}{20}    a \,  b_1^2   
+ \frac{11}{28 }    a \, b_2^2  
\bigg) Y_{10}
\Bigg].
\end{split}
\end{equation}

Now turning to the right hand side of the equation, 
let us note first that
\begin{equation}\label{eq:edth2edth2-1}
\bar\eth^2 V_2  \eth^2 V_2 =
6 b_2^2 
\Big(
\frac{1}{\sqrt{4\pi}} Y_{00}
- \frac{1}{7} \sqrt{\frac{5}{ \pi }} Y_{20}
+ \frac{1}{7\sqrt{4 \pi }}   Y_{40}
\Big) ;
\end{equation}
so that the $l=0$ part of expression (\ref{eq:edth2edth2}) is
\begin{equation}\label{eq:edth2edth2-2}
\hat \gamma^2   
6 b_2^2 
\Big(
\frac{1}{\sqrt{4\pi}} Y_{00}
\Big)
;
\end{equation}
while the $l=1$ part of (\ref{eq:v1edth2edth2}) becomes
\begin{equation}\label{eq:v1edth2edth2-2}
\begin{split}
- \hat \gamma^3 V_1 & \bar\eth^2 V_2  \eth^2 V_2 
= 
- \hat \gamma^3 b_1 b_2^2 \frac{9}{14\pi} Y_{10}
+ \text{(other terms)}
 .
\end{split}
\end{equation}

So that
\begin{equation}
\mathbf{F}_0^0 
=
- 
\hat \gamma^2   
b_2^2 6 
\Big(
\frac{1}{\sqrt{4\pi}} Y_{00}
\Big)
,
\end{equation}
and
\begin{equation}
3 \hat{\mathbf{F}}_0^{(3)}
=
- \hat \gamma^3 b_1 b_2^2 \frac{9}{14\pi} Y_{10}
.
\end{equation}

Then, the complete equations are:
\begin{equation}
\begin{split}\label{eq:lfh-rhs-l0}
3 M 
&
\hat\gamma \big(
- 3   \frac{a b_1}{\sqrt{4 \pi}}
+ \dot V_0 + \mathscr{T}_0(\dot V_1 )
\big) 
Y_{00}  \\
=&
\hat \gamma^2   
b_2^2 6 
\Big(
\frac{1}{\sqrt{4\pi}} Y_{00}
\Big)
;
\end{split}
\end{equation}
and
\begin{equation}
\begin{split}\label{eq:lfh-rhs-l1}
3 M 
\Bigg(&
a Y_{10}
+
\hat\gamma \Big(
- 3 \big( \frac{a b_0}{\sqrt{4 \pi}} + \frac{a b_2}{\sqrt{5 \pi}} \big) Y_{10} 
+ \mathscr{T}_0(\dot V_1 )
\Big) \\
+&
\hat{\gamma}^2 6
\frac{1}{\pi} 
\bigg(
\frac{1}{4} 
a \,b_0^2  + \frac{1}{\sqrt{5}} a \, b_0 b_2    \\
+& a \,  b_1^2   \big(\frac{1}{4}  + \frac{1}{5}
\big)  
+ a \, b_2^2  
\big(\frac{1}{4  }  +  \frac{1}{7} 
\big)  
\bigg) Y_{10}
\Bigg) \\
=&
-\Big( \hat \gamma^3 b_1 b_2^2 \frac{9}{14\pi} Y_{10} \Big)
\big( 1 +  \frac{1}{ \Upsilon } \big)
;
\end{split}
\end{equation}
where one notices that this equation forces us to make a correction
to the first order equations of motion, so that the $\dot V_1 $
is balanced. We will clarify this below.

Let us express (\ref{eq:lfh-rhs-l0}) by
\begin{equation}
\begin{split}\label{eq:lfh-rhs-l0-clean}
\frac{3}{2} M 
\hat\gamma \big(
-   a b_1
+ \sqrt{4\pi} \big( \dot V_0 + \mathscr{T}_0(\dot V_1 ) \big)
\big)   
=
\hat \gamma^2  b_2^2 
.
\end{split}
\end{equation}

Let us express this proportionality between $b_1$ and $a$
 by defining $\alpha$ to be
the factor
\begin{equation}\label{eq:b1}
\hat\gamma \, b_1 = - \alpha M a .
\end{equation}
so that the quantity $\alpha$ has no units and, in first order
of back reaction effects, it is equivalent to say that
\begin{equation}
\hat\gamma \, v_1^\mu = - \alpha M \mathbf{a}^\mu ;
\end{equation}
that is,
although $v_1^\mu$ was actually defined in terms of the 
spacelike part of $\mathbf{f}^\mu$, since the difference 
between $\mathbf{a}^\mu$ and $\mathbf{f}^\mu$, is order $\hat \gamma$,
one will have that expressing $\hat \gamma v_1^\mu$
in terms of $\mathbf{a}^\mu$, instead of $\mathbf{f}^\mu$, will
introduce differences of order $\hat \gamma^2$.

Then we have that
\begin{equation}\label{eq:lfh-rhs-l0-clean-2}
\begin{split}
 \frac{3}{2} \alpha M^2 a^2  
 +
 \frac{3}{2} \sqrt{4\pi} M  \hat\gamma \big( \dot V_0 + \mathscr{T}_0(\dot V_1 ) \big)
= 
\hat \gamma^2   \, b_2^2 
.
\end{split}
\end{equation}

Let us study in more detail the term involving $\dot  V_1$.
If we use the four dimensional notation, then one would say
that 
\begin{equation}
\hat\gamma \, \dot v_1^\mu = - \dot\alpha M  \mathbf{a}^\mu 
- \alpha M  \dot{\mathbf{a}}^\mu 
;
\end{equation}
and so, in terms of the scalar, we have
\begin{equation}
\begin{split}
\hat\gamma \, \dot V_1 =
- \dot\alpha M  \mathcal{A} - \alpha M  \dot{\mathcal{A}}
;
\end{split}
\end{equation}
with
\begin{equation}
\dot{\mathcal{A}} = 
\dot{\mathbf{a}}^\mu \frac{\hat l_\mu}{V_\eta}
-  \mathcal{A}^2 ;
\end{equation}
since $\hat\gamma \, V_1 = - \alpha M \mathcal{A}$.

The $l=0$ term of $\dot V_1$, in the rest frame, is
\begin{equation}
\begin{split}
\hat\gamma \, \mathscr{T}_0(\dot V_1 ) =&
- \alpha M  \mathscr{T}_0(\dot{\mathcal{A}} )
=
- \alpha M  \big(
\dot{\mathbf{a}}^0
- \mathscr{T}_0( \mathcal{A}^2 ) \big) \\
=&
\frac{2}{3} \alpha M  \mathbf{a}^\mu  \mathbf{a}_\mu 
;
\end{split}
\end{equation}
where it should be remarked that $\mathbf{a}^\mu  \mathbf{a}_\mu \leqslant 0$.

Then, in order to consistently consider the term including the
symbol $\dot V_1$ we should increase the order of the 
balanced equations of motion; so that we generalize
(\ref{eq:mot-compact}) to
\begin{equation}\label{eq:mot-compact-o1}
\boxed{
\mathbf{a}^\mu =  \mathbf{f}^\mu 
 + \dot{\big( \alpha M  \mathbf{f}^\nu \big)}  
 + \beta M \mathbf{f}^\nu \mathbf{f}_\nu \mathbf{v}^\mu
+ f_\lambda^\mu ;
}
\end{equation}
where $\dot{\big( \alpha M  \mathbf{f}^\nu \big)}$ denotes the time
derivative of ${\big( \alpha M  \mathbf{f}^\nu \big)}$.
This constitutes our second order version of the balanced
equations of motion; while (\ref{eq:mot-compact}) was the first order
version.
Note that now we will have a modification of the $l=1$ term of the
field equation that will balance the spacelike part of $\dot V_1$.
Instead of (\ref{eq:AigFmasflambda}) we now have
\begin{equation}\label{eq:AigFmasflambda-1}
\mathcal{A} = \mathcal{F} 
+   \dot{\big( \alpha M  \mathbf{f}^\nu \big)}  l_\nu
+ \beta M \mathbf{f}^\nu \mathbf{f}_\nu
+ f_\lambda .
\end{equation}
Then, instead of (\ref{eq:vacuum2-fieldfh-con-V0})  we have
\begin{equation}\label{eq:field-f0-5}
\begin{split}
3 M &
\bigg(
\Big(- \hat{\mathcal{F}}^{(3)}
+ (\beta - \frac{1}{3}\alpha) M \mathbf{f}^\nu \mathbf{f}_\nu
\Big)
\\
&+
\hat\gamma \big[
- 3 \Big(- \hat{\mathcal{F}}^{(3)}
\Big)  (V_0 + V_1 + V_2) 
+
\dot V_0 
\big] \\
&+
\hat\gamma^2 
6 \Big(- \hat{\mathcal{F}}^{(3)}
\Big) (V_0 + V_1 + V_2)^2  + \mathscr{O}(\hat\gamma^3)
\bigg) \\
=&
- \mathbf{F}_0^0 
+ 3 \, \hat{\mathbf{F}}_0^{(3)}  \big( 1 +  \frac{1}{ \Upsilon } \big)
+ \text{terms with }\hat Y_{l_2,m} . 
\end{split}
\end{equation}

The $l=0$ term of this relation is
\begin{equation}\nonumber
\begin{split}
3 M 
\bigg(
&
 (\beta - \frac{1}{3}\alpha) M \mathbf{f}^\nu \mathbf{f}_\nu
+
\hat\gamma \big(
- 3   \frac{a b_1}{\sqrt{4 \pi}} Y_{00}  
+ \dot b_0 Y_{00} 
\big) 
\bigg)
\\
= &
\hat \gamma^2  b_2^2 \Big(
\frac{6}{\sqrt{4\pi}} Y_{00}
\Big)
.
\end{split}
\end{equation}
Replacing $b_1$ we have
\begin{equation}\label{eq:lfh-rhs-l0-clean-b}
\boxed{
\begin{split}
M 
\bigg(
&
4\pi 
(\beta - \frac{1}{3}\alpha) M \mathbf{f}^\nu \mathbf{f}_\nu
+ 3 \alpha M  a^2
+ \hat\gamma \dot b_0 \sqrt{4\pi}
\bigg)
\\
= &
2 \, \hat \gamma^2  b_2^2 
;
\end{split}
}
\end{equation}

Recall that in the frame in which the acceleration is in the $z$
direction $a = -a^z \sqrt{\frac{4\pi}{3}}$ and
$\mathbf{f}^\nu \mathbf{f}_\nu = -(a^z)^2$;
so that we could also express it as
\begin{equation}\label{eq:lfh-rhs-l0-clean-d}
M 
\bigg(
2\pi M (a^z)^2
\big(\frac{4}{3}\alpha - \beta \big)
+ \hat \gamma \dot b_0 \sqrt{\pi}
\bigg)
= 
\hat \gamma^2  b_2^2 
.
\end{equation}

The $l=1$ component of the equation now is
\begin{equation}\label{eq:lfh-rhs-l1-2b}
\begin{split}
M 
\Bigg(&
a 
+
\hat\gamma \Big(
- 3 ( \frac{a b_0}{\sqrt{4 \pi}} + \frac{a b_2}{\sqrt{5 \pi}} )
\Big) \\
+&
\hat{\gamma}^2 
\frac{6}{\pi} 
\bigg(
\frac{1}{4} 
a \,b_0^2  + \frac{1}{\sqrt{5}} a \, b_0 b_2    \\
+& \frac{9}{20} a \,  b_1^2  
+ \frac{11}{28} a \, b_2^2  
\bigg) 
\Bigg) \\
=&
-\frac{3}{14\pi}  \hat \gamma^3 b_1 b_2^2 
\big( 1 +  \frac{1}{ \Upsilon } \big) 
.
\end{split}
\end{equation}
and replacing $b_1$ we have
\begin{equation}\label{eq:lfh-rhs-l1-2}
\boxed{
\begin{split}
M 
\Bigg(&
a 
+
\hat\gamma \Big(
- 3 ( \frac{a b_0}{\sqrt{4 \pi}} + \frac{a b_2}{\sqrt{5 \pi}} )
\Big) \\
+&
\hat{\gamma}^2 6
\frac{1}{\pi} 
\bigg(
\frac{1}{4} 
a \,b_0^2  + \frac{1}{\sqrt{5}} a \, b_0 b_2    \\
+& 
\frac{11}{28}  a \, b_2^2  
\bigg)
+
\frac{27}{10\pi} a \,  (\alpha M a)^2  
\Bigg) \\
=&
\frac{3}{14\pi}  \alpha M a \Big( \hat \gamma^2 b_2^2  \Big)
\big( 1 +  \frac{1}{ \Upsilon } \big) 
.
\end{split}
}
\end{equation}

Summarizing, the balanced equations of motion 
at second order is given by (\ref{eq:mot-compact-o1})
where $\alpha$, $\beta$ and $b_1$ and $b_2$ appearing in $f_\lambda^\mu$, satisfy
equations (\ref{eq:b1})
(\ref{eq:lfh-rhs-l0-clean-b}) and (\ref{eq:lfh-rhs-l1-2}).
Comparing with the electromagnetic case, one would be tempted to take
$\beta = \alpha$; but we leave the setting of $\beta$ open for
future work.

This is our main result.
The tight connexion between the interior and asymptotic structure
in the null gauge model for compact objects, has led us
to extend the study to the second order balanced equations of motion,
as expressed by (\ref{eq:mot-compact-o1}).

We will study how to fix these remaining degrees of freedom
by applying these equations to specific observations;
which will be carried out in future work.

\section{Final comments}

We have applied the general framework for the construction of balanced equations
of motion introduced in \cite{Gallo:2016hpy} to the particular case of
the null gauge, that we have described here.

We have presented the balanced equations of motion which are coupled
to the field equation of the monopole geometry.
We have indicated the way in which system (\ref{eq:vacuum1-b}) can be solved numerically
in an exact form; and we have also presented an expansion of the field
equation in terms of orders.
From the tight relations between the interior and asymptotic structure
in the null gauge model, we have arrived at the second order
balanced equations of motion  (\ref{eq:mot-compact-o1});
where the intervening quantities $b_0$, $\alpha$ and $b_2$
are related by the components of the field equation
(\ref{eq:lfh-rhs-l0-clean-b}) and (\ref{eq:lfh-rhs-l1-2}).

After having deduced the form of the balanced equations of motion
at second order in the acceleration, the idea is that
if one tackles system (\ref{eq:vacuum1-b}) by numerical method,
one would obtain higher accuracy using the second order version
of the equations of motion.

Since we do not request low velocities, or weak fields,
or condition on the masses, 
we expect with our models to improve on the range of possible
systems that we can study with respect to those covered by the
post-Newtonian and the self-force approaches.

When applying these balanced equations of motion to a binary
system of comparable masses, 
one must deal with two dynamical times; one for each particle.
So the problem turns into a bound binary retarded dynamical system.
When dealing with this type of retarded system, it is common
to recur to approximations using some kind of universal dynamical
time; as is the case in the post-Newtonian approaches.
Instead we intend to apply a method we have developed in\cite{Moreschi:2018gmf},
based on high order approximations of the trajectory
from the force equations.

In the process of relating the equations of motion to the
monopole field equation, we have solve the $l=0$ and $l=1$ terms
of the field equation at the $G^2$ and $G^3$ order respectively.
This is what is needed to take into account the first effects
of back reaction to the equations of motion in the null gauge.

A property of the null gauge model is that one can request
increasing precision in the model by demanding
other components of the field equations to be satisfied;
i.e. $l=2,3,...$, which of course will require the 
introduction of new quantities, as $b_3$, $b_4$, etc.
At any stage, the neglected $b_i$ are assumed to be less
important, since they represent internal degrees of freedom
of the compact object.
But there might be occasions in which one would like to 
resort to this internal structure; as for example the case
to build a model of non-black holes compact objects,
as are the neutron stars.
The degree in which this type of model can be successful
for the description of these systems will be a matter of
forthcoming work.

The first method that we have suggested for solving the model,
is to solve numerically 
the coupled system of the balanced equation of motion
and the exact monopole field equation (\ref{eq:vacuum1-b}) numerically;
which, it should be emphasized, it provides with a global exact solution
of the monopole metric, that is, for all $r$.
This, we think, is a very nice property of the null gauge approach
to the dynamics of the binary system.
So, it seem that
the tight relation between the interior and asymptotic
structure of the null gauge model, along null directions,
of this gauge, will provide more physically interesting
results than those from the model based on the harmonic gauge
presented in \cite{Gallo:2017yys}.
All these issues will become clear in the application of our models
to observations of gravitational waves.

\appendix

\section{Appendix}

\subsection{Coordinate systems and the edth operator}
Normally the edth operator is expressed in terms of the complex coordinates 
$(\zeta, \bar\zeta)$ or in terms of the spherical coordinates $(\theta,\phi)$.
The relation between the two coordinate systems is given by
\begin{equation}\label{eq:zetadetetafi}
 \zeta = e^{i \phi} \cot(\frac{\theta}{2}) .
\end{equation}
From this one deduces that the coordinate complex vector is:
\begin{equation}\label{eq:dzeta}
 \frac{\partial}{\partial \zeta} = - e^{-i \phi} \sin^2(\frac{\theta}{2}) 
\Big(
 \frac{\partial}{\partial \theta} + \frac{i}{\sin(\theta)}\frac{\partial}{\partial \phi}
\Big) .
\end{equation}

In the unit sphere, the complex vectors $m$ and $\bar m$ are chosen so that
$m \propto \frac{\partial}{\partial \zeta}$ and therefore
$\bar m \propto \frac{\partial}{\partial \bar\zeta}$.
Defining
\begin{equation}
 P_0 = \frac{1}{2} \big( 1 + \zeta \bar\zeta \big) ;
\end{equation}
one writes, for the unit sphere, 
\begin{equation}
 m = \sqrt{2} P_0 \frac{\partial}{\partial \zeta} ,
\end{equation}
and
\begin{equation}
 \bar m = \sqrt{2} P_0 \frac{\partial}{\partial \bar\zeta} .
\end{equation}
Note that since the scalar product between these vectors is minus one, we
have
\begin{equation}
 m_a \Leftrightarrow - \frac{d \bar\zeta}{\sqrt{2} P_0} .
\end{equation}
Then, according with our signature conventions one has
\begin{equation}
 -m_a \bar m_b - m_b \bar m_a 
 \Leftrightarrow 
- \frac{d \zeta \, d \bar\zeta}{P_0^2} 
= - \Big( d\theta^2 + \sin^2(\theta) d\phi^2 \Big) .
\end{equation}

Whenever needed we also use the decomposition of the complex coordinates
in terms of
\begin{equation}
 \zeta = \frac{1}{2} ( x^2 + i x^3) ,
\end{equation}
and similarly
\begin{equation}
 \bar\zeta = \frac{1}{2} ( x^2 - i x^3) ;
\end{equation}
so that the coordinate vectors are:
\begin{equation}
 \frac{\partial}{\partial \zeta}
= \frac{\partial}{\partial x^2} - i \frac{\partial}{\partial x^3} ,
\end{equation}
and
\begin{equation}
 \frac{\partial}{\partial \bar\zeta}
= \frac{\partial}{\partial x^2} + i \frac{\partial}{\partial x^3} .
\end{equation}

Using now the GHP\cite{Geroch73} definition of the edth operator,
applied to the unit sphere one can write
\begin{equation}
  \label{eq:ethoper0}
  \eth f = \sqrt{2} P_0^{1-s} \frac{\partial}{\partial \zeta}\left(P_0^s f\right)
\end{equation}
and 
\begin{equation}
  \label{eq:ethboper0}
  \bar\eth f = \sqrt{2} P_0^{1+s} \frac{\partial}{\partial \bar\zeta}
  \left(P_0^{-s} f\right) ;
\end{equation}
where $s$ is the spin weight of the quantity $f$.

Comparing the GHP expressions with equation (3.9) of \cite{Newman66}
we see that
\begin{equation}
 \eth = \frac{1}{\sqrt{2}} \eth_P ;
\end{equation}
where $\eth_P$ is the NP definition.
Instead, their equation (3.8) would need an extra factor $e^{-i \phi}$,
as it can be deduced from (\ref{eq:dzeta}); since they use a rotated
null tetrad.

\subsection{The edth operator as intrinsic objects on the spheres
and the spin $s$ spherical harmonics}

In general for a sphere with metric $d\zeta d\bar \zeta /P^2$, and $P=P_0 V$,
 one can define the intrinsic edth operator 
acting on a function $f$ of spin weight $s$ by
\begin{equation}
\eth_V f = \sqrt{2} P^{1-s} \frac{\partial(P^s f)}{\partial \zeta}
\end{equation}
and similarly
\begin{equation}
\bar\eth_V f = \sqrt{2} P^{1+s} \frac{\partial(P^{-s} f)}{\partial \bar\zeta}.
\end{equation}
The commutator of these two operators is
\begin{equation}
\left(
  \bar\eth_V \eth_V - \eth_V \bar\eth_V
\right)
f = s \, K_V  \,f ;
\end{equation}
where it is important to notice that the convention that comes from the
GHP\cite{Geroch73} formalism, differs from the original one
suggested by Newman and Penrose in reference \cite{Newman66}.
When $P=P_0$, equivalently $V_M=1$, it is convenient to refer to
the spin $s$ spherical harmonics\cite{Newman66} 
${}_sY_{lm}$ which have the following
properties
\begin{equation}
\eth \,{}_sY_{lm} = \sqrt{\frac{(l-s)(l+s+1)}{2}}\; {}_{s+1}Y_{lm} ,
\end{equation}
\begin{equation}
\bar\eth \,{}_sY_{lm} = -\sqrt{\frac{(l+s)(l-s+1)}{2}}\; {}_{s-1}Y_{lm} 
\end{equation}
and
\begin{equation}
\bar\eth \eth \,{}_sY_{lm} = -\frac{(l-s)(l+s+1)}{2}\; {}_{s}Y_{lm} .
\end{equation}
This last eigenvalue problem is useful in classifying the functions
on the sphere with metric  $d\zeta d\bar \zeta /(V^2_M P_0^2)$
even when $V_M \neq 1$.

\subsection{Product of spherical harmonics and spin-weighted spherical harmonics}\label{ap:sph-harmo}

In this section to help the reading we will use the notation
$Y_l^m \equiv Y_{lm}$.

It is convenient to recall
that\footnote{(16.89) Merzbacher\cite{Merzbacher70}}
the product of two spherical harmonics can be express, in terms of
Clebsch-Gordan coefficients by
{\small 
\begin{equation}\label{eq:prodyl1poryl2}
\begin{split}
 Y_{l_1}^{m_1} & (\theta,\phi) Y_{l_2}^{m_2}(\theta,\phi) 
=
\sum_l \sqrt{\frac{(2 l_1 + 1)(2 l_2 + 1)}{4 \pi (2l+1)}}
< l_1 l_2 0 0 | l_1 l_2 l 0 > \\
& \quad < l_1 l_2 m_1 m_2 | l_1 l_2 l  (m_1+m_2) >
Y_{l}^{m_1 + m_2} (\theta,\phi) . 
\end{split}
\end{equation}
}

For the product of two spin-weighted spherical harmonics one has\cite{Held70}
{\small 
\begin{equation}\label{eq:prods1yl1pord2yl2}
\begin{split}
_{s_1}Y_{l_1}^{m_1} (\theta,\phi) &\; _{s_2}Y_{l_2}^{m_2}(\theta,\phi) \\
=&
\sum_l \sqrt{\frac{(2 l_1 + 1)(2 l_2 + 1)}{4 \pi (2l+1)}}
< l_1, l_2 ; -s_1 , -s_2 | l , -s > \\
& \quad < l_1 , l_2 ; m_1 , m_2 | l , m > \, 
_{s}Y_{l}^{m} (\theta,\phi) ;
\end{split}
\end{equation}
}
with $m = m_1+m_2$, $s = s_1+s_2$ and $|l_1-l_2| \leqslant l \leqslant |l_1+l_2|$,
and with a slight change in the notation of the Clebsch-Gordan coefficients.

Then, in particular one has
\begin{equation}\label{eq:prods1yl1pord2yl2-2020}
\begin{split}
_{2}Y_{2}^{0} (\theta,\phi) &\; _{-2}Y_{2}^{0}(\theta,\phi) \\
=&
\sum_l \sqrt{\frac{5^2}{4 \pi (2l+1)}}
< 2 , 2 , -2 , 2 | l , 0 > \\
& \quad < 2, 2, 0, 0 | l , 0 > \, 
Y_{l}^{0} (\theta,\phi) \\
=&
\frac{5}{\sqrt{4 \pi }}
\Big(
\frac{1}{5} Y_0^0
- \frac{2}{7\sqrt{5}} Y_2^0
+ \frac{1}{35} Y_4^0
\Big) \\
=&
\frac{1}{\sqrt{4\pi}} Y_0^0
- \frac{1}{7} \sqrt{\frac{5}{ \pi }} Y_2^0
+ \frac{1}{7\sqrt{4 \pi }}   Y_4^0
.
\end{split}
\end{equation}

\subsection*{Acknowledgments}

We acknowledge support from CONICET, SeCyT-UNC and Foncyt.

%
%

\end{document}